\documentclass[12pt]{article}
\let\section=\subsection     \let\subsection=\subsubsection                %%
\usepackage{graphicx}

\begin{document}
\begin{center}
   {\large \bf The astrophysical rate of $^{15}$O($\alpha$,$\gamma$)$^{19}$Ne via recoil-decay tagging and its implications for nova nucleosynthesis}\\[5mm]
B. Davids\\[5mm]
{\small \it  Kernfysisch Versneller Instituut, Zernikelaan 25, 9747 AA Groningen, \\The Netherlands}\\[5mm]
J. Jos\'e\\[5mm]
{\small \it Dept. F\'{\i}sica i Enginyeria Nuclear, Univ. Polit\'ecnica de Catalunya, and \\
Institut d'Estudis Espacials de Catalunya, Ed. Nexus-201, Gran Capit\`a 2-4, 08034 Barcelona, Spain} 
\end{center}

\begin{abstract}\noindent
The $^{15}$O($\alpha$,$\gamma$)$^{19}$Ne reaction is one of two known routes for breakout from the hot CNO cycles into the $rp$ process. Its astrophysical rate depends on the decay properties of excited states in $^{19}$Ne lying just above the $^{15}$O + $\alpha$ threshold. We have measured the $\alpha$-decay branching ratios for these states using the $p(^{21}$Ne,$t)^{19}$Ne reaction at 43 MeV/u. Combining our branching ratio measurements with previous determinations of the radiative widths of these states, we calculate the astrophysical rate of $^{15}$O($\alpha,\gamma)^{19}$Ne. Using this reaction rate, we perform hydrodynamic calculations of nova outbursts and conclude that no significant breakout from the hot CNO cycles into the $rp$ process occurs in classical novae via $^{15}$O($\alpha,\gamma)^{19}$Ne.
\end{abstract}

Novae are thermonuclear runaways powered by the accretion of hydrogen-rich material from a stellar companion onto the surface of a white dwarf in a close binary system. Energy production and nucleosynthesis in such stellar sites are in general determined by the CNO cycles, with contributions from the NeNa and MgAl cycles \cite{jose98}. 
The possibility of breakout from the hot CNO cycles into the $rp$ process has been extensively discussed in previous work, in particular under the
high temperature and density conditions attained in the most massive ONe nova outbursts. Several reactions have been suggested as pathways for this breakout 
\cite{wiescher99}, but only two are currently thought to be possibilities: $^{15}$O($\alpha$,$\gamma$)$^{19}$Ne and $^{18}$Ne($\alpha$,$p$)$^{21}$Na. In 
astrophysical environments the $^{15}$O($\alpha$,$\gamma$)$^{19}$Ne reaction proceeds predominantly through resonances lying just above the $^{15}$O + $\alpha$ threshold at 3.529 MeV in $^{19}$Ne. For nova conditions in particular, the reaction rate is determined by the $\alpha$ width $\Gamma_{\alpha}$ of the 4.033 MeV, 3/2$^+$ state, owing both to its close proximity to the $^{15}$O + $\alpha$ threshold and its low centrifugal barrier to $\alpha$ capture.

Direct measurements of the low energy cross section, which require high-intensity radioactive $^{15}$O beams, are planned. For states in $^{19}$Ne lying at excitation energies relevant to novae and accreting neutron stars, only the $\alpha$- and $\gamma$-decay channels are open, as the proton and neutron separation energies are 6.4 and 11.6 MeV \cite{audi95} respectively. Hence, by populating these states and observing the subsequent $\alpha$ and $\gamma$ decays, one can deduce the branching ratio B$_{\alpha}\equiv\Gamma_{\alpha}/\Gamma$. If $\Gamma_{\gamma}$ is also known, one can then calculate $\Gamma_{\alpha}$ and thereby the contribution of each state to the resonant rate of $^{15}$O($\alpha$,$\gamma$)$^{19}$Ne. Efforts of this kind to detect $\alpha$ particles from the decay of $^{19}$Ne states populated via transfer reactions were made \cite{magnus90,laird02}, but these experiments were insufficiently sensitive to measure B$_{\alpha}$ for the critical 4.033 MeV state, which was expected to be of order 10$^{-4}$ \cite{langanke86}.

In an experiment at the Kernfysisch Versneller Instituut \cite{davids03}, we have succeeded in obtaining branching ratio data at this level of sensitivity. Populating the important states via the $^{21}$Ne($p,t)^{19}$Ne reaction in inverse kinematics with a $^{21}$Ne beam energy of 43 MeV/u, we detected either $^{19}$Ne recoils or their $^{15}$O $\alpha$-decay products in coincidence with tritons in the Big-Bite Spectrometer (BBS) \cite{berg95}. The large momentum acceptance of the BBS ($\Delta$p/p~=~19\%) allowed detection of either $^{19}$Ne recoils or $^{15}$O decay products along with tritons emitted backward in the center of mass system. Positioning the BBS at 0$^{\circ}$ maximized the yield to the 4.033 MeV, 3/2$^+$ state in $^{19}$Ne. This state, whose dominant shell-model configuration is $(sd)^5 (1p)^{-2}$ \cite{fortune78}, was selectively populated by an $\ell=0$, two-neutron transfer from the 3/2$^+$ ground state of $^{21}$Ne. Position measurements in two vertical drift chambers (VDCs) \cite{woertche01} allowed reconstruction of the triton trajectories. Excitation energies of the $^{19}$Ne residues were determined from the kinetic energies and scattering angles of the triton ejectiles. The $\gamma$ decays of states in $^{19}$Ne were observed as $^{19}$Ne-triton coincidences in the BBS, whereas $\alpha$ decays were identified from $^{15}$O-triton coincidences.

Recoils and decay products were detected and stopped just in front of the VDCs by fast-plastic/slow-plastic phoswich detectors \cite{leegte92} that provided energy loss and total energy information. A separate array of phoswich detectors was used to identify tritons after they passed through the VDCs. Timing relative to the cyclotron radio frequency signal was also employed for unambiguous particle identification. An Al plate prevented many of the heavy ions copiously produced by projectile fragmentation reactions of the $^{21}$Ne beam in the (CH$_2$)$_n$ target from reaching the VDCs. The spatial extent of the heavy-ion phoswich array was sufficient to guarantee 100\% geometric efficiency for detection of $^{19}$Ne recoils and $^{15}$O decay products for $^{19}$Ne excitation energies~$\leq$~5.5 MeV. This resulted largely from the forward focusing of the $^{19}$Ne recoils, which emerged at angles~$\leq$~0.36$^{\circ}$ for tritons with scattering angles~$\leq$~4$^{\circ}$. The low decay energies of the states studied limited the angular and energy spreads of the $^{15}$O decay products.

The $^{19}$Ne excitation energy spectrum obtained from $^{19}$Ne-triton coincidences, representing $\gamma$ decays of states in $^{19}$Ne, is shown in Fig.\ 1. Its most prominent peak is due to the 4.033 MeV, 3/2$^+$ state. The 4.379 MeV, 7/2$^+$ state is the second most strongly populated. Contributions from other known states \cite{tilley95} are indicated. The experimental resolution of 90 keV FWHM is insufficient to resolve the 4.140 and 4.197 MeV states from one another; the 4.549 and 4.600 MeV states are also unresolved. However, the astrophysically important 4.033 and 4.379 MeV states are well separated from the others. The curve shown in Fig.\ 1 is the sum of a constant background and 8 Gaussians centered at the known energies of the states, with standard deviations fixed by the experimental resolution of 90 keV FWHM.

\begin{figure}\begin{center}\includegraphics[width=8cm]{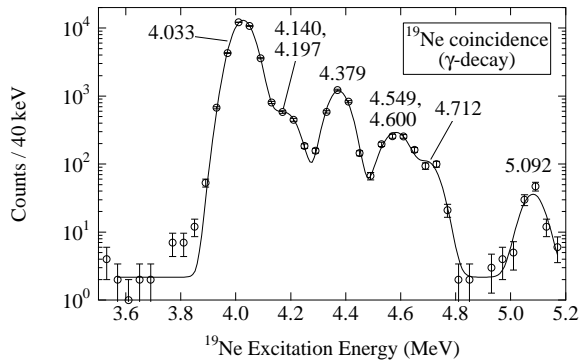} \caption{$^{19}$Ne-triton
coincidences ($\gamma$ decays of states in $^{19}$Ne). The curve is the sum of a constant background and 8 Gaussians centered at the energies of known states in $^{19}$Ne, the widths of which were determined by the experimental resolution of 90 keV FWHM.} \end{center}\end{figure}

Fig.\ 2 shows the $^{19}$Ne excitation energy spectrum obtained from $^{15}$O-triton coincidences, corresponding to $\alpha$ decays of states in $^{19}$Ne. The observed states are labeled by their energies; the 4.549 and 4.600 MeV states cannot be resolved. A fit consisting of Gaussians plus a constant background is shown as well. The states below 4.549 MeV decay overwhelmingly by $\gamma$ emission, while the higher-lying states observed here decay preferentially by $\alpha$ emission, simply because the larger available decay energy enhances the barrier penetrability. The background represents a larger fraction of the total events in the $^{15}$O-triton coincidence spectrum than in the $^{19}$Ne-triton coincidence spectrum, though it is still very small.  This background is well reproduced by a constant, the value of which is determined to a precision of 7\% (1$\sigma$).

\begin{figure}\begin{center}\includegraphics[width=8cm]{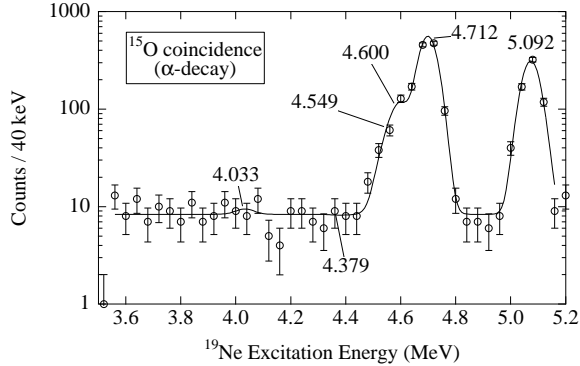} \caption{$^{15}$O-triton
coincidences ($\alpha$ decays of states in $^{19}$Ne). The curve is the sum of a constant background and 6 Gaussians corresponding to known states in $^{19}$Ne, the widths of which were fixed by the experimental resolution. The $^{15}$O + $\alpha$ threshold lies at 3.529 MeV.} \end{center}\end{figure}

\begin{figure}\begin{center}\includegraphics[width=13cm]{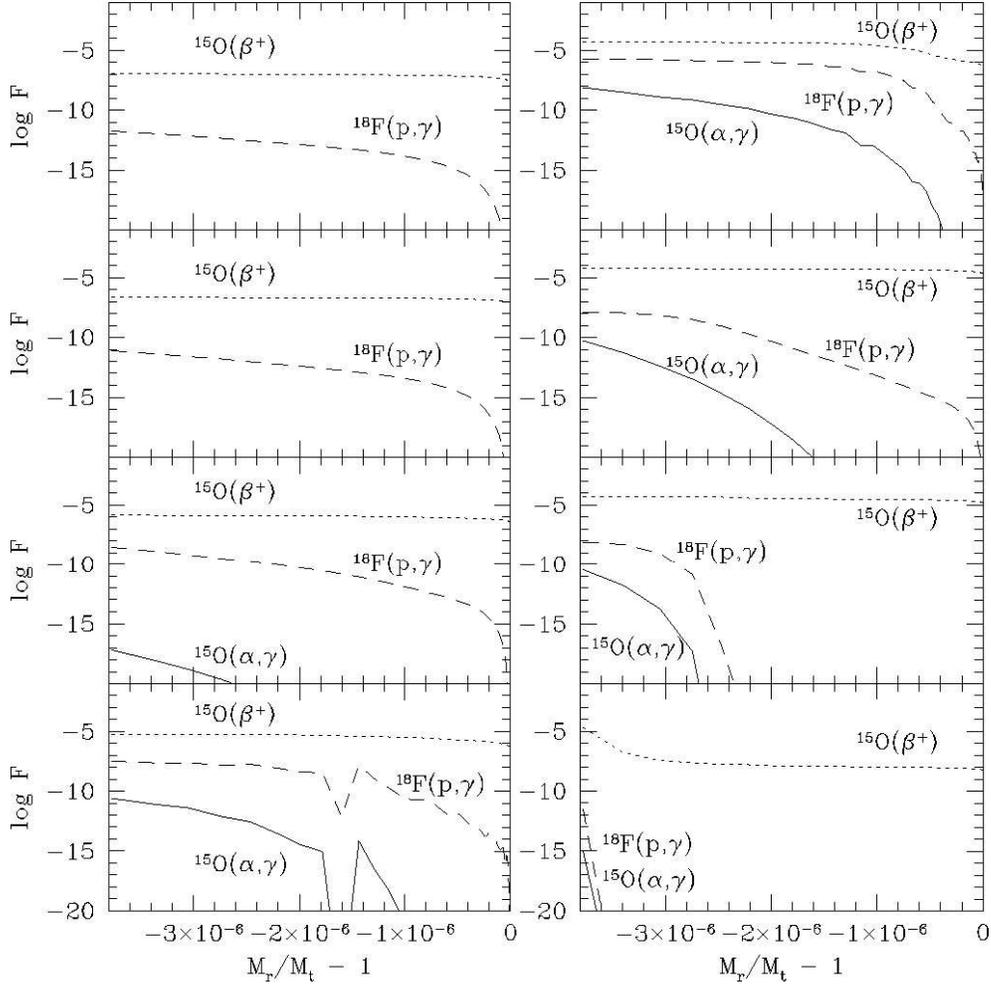} \caption{A sample of relevant reaction fluxes F (s$^{-1}$) in logarithmic scale along the accreted envelope,
for a 1.35 solar mass ONe nova accreting at a rate $\dot M = 2 \times 10^{-10}$ $M_\odot$ yr$^{-1}$. The mass coordinate represents the mass below the surface relative to the total mass. From top to bottom, panels correspond to a time series from the early stages of the explosion up to the ejection stage, with a temperature at the burning shell equal to: $9 \times 10^7$, $10^8$, $1.5 \times 10^8$, $2.5 \times 10^8$, and $T_{peak} = 3.21 \times 10^8$ K (top right panel), plus three panels corresponding to the last phases of the explosion, when the white dwarf envelope has expanded to a size of $R_{wd} \sim 10^9, 10^{10}$, and $10^{12}$ cm, respectively. The base of the ejected shells lies at a mass coordinate of $-3.05 \times 10^{-6}$.}\end{center}\end{figure}

From these excitation energy spectra, we determine the $\alpha$-decay branching ratios of six states in  $^{19}$Ne. No statistically significant evidence for $\alpha$ decays from the states at 4.033 and 4.379 MeV was observed. For these states the $\alpha$- and $\gamma$-decay spectra were numerically integrated in 100 keV intervals centered at the known energies of the states. These data were subjected to both Bayesian and classical statistical analyses to determine upper limits on the $\alpha$-decay branching ratios at various confidence levels. The more conservative Bayesian analysis yields a 90\% confidence level upper limit on B$_{\alpha}$ for the 4.033 MeV state of $4.3\times10^{-4}$. Using independently determined radiative widths for these states, e.g. \cite{hackman00}, we compute their $\alpha$ widths and a very conservative 99.73\% confidence level upper limit on the astrophysical rate of $^{15}$O($\alpha$,$\gamma$)$^{19}$Ne. Details of this procedure can be found in Ref.\ \cite{davids03}.

With this $^{15}$O($\alpha$,$\gamma$)$^{19}$Ne rate, we have performed a series of hydrodynamic calculations of nova outbursts on 1.35 M$_\odot$ ONe white dwarfs using an implicit, spherically symmetric, hydrodynamic code in Lagrangian formulation \cite{jose98}. This code describes the entire outburst from the onset of accretion through the thermonuclear runaway, including the expansion and ejection of the accreted envelope. We assume a mass accretion rate of $2\times10^{-10}$ M$_\odot$ yr$^{-1}$ and 50\% mixing between 
the solar-like accreted material and the outermost, ONe-rich shells of the underlying white dwarf.  As shown in Fig.\ 3, confirming the conclusions of Refs.\ \cite{davids03,iliadis02},  
 the $^{15}$O($\alpha$,$\gamma$)$^{19}$Ne rate is less than 0.02\% of the $^{15}$O $\beta^+$ decay rate throughout the whole envelope, including its hottest, innermost shell
(i.e., the base of the envelope),  which is not ejected by the explosion. 
 Since the timescale of the outburst is only a few times the timescale of the hot CNO cycles \cite{davids03}, this implies that only trace amounts of CNO material are processed via $^{15}$O($\alpha$,$\gamma$)$^{19}$Ne, even in the material retained by the white dwarf after ejection of the accreted envelope. Moreover, as Fig.\ 3 illustrates, the contribution of $^{15}$O($\alpha$,$\gamma$) to the production of $^{19}$Ne is greatly exceeded by the contribution of $^{18}$F(p,$\gamma$). In conclusion, no significant enrichment of nova ejecta due to hot CNO breakout via $^{15}$O($\alpha$,$\gamma$)$^{19}$Ne is expected.

This work was performed as part of the research program of the {\it Stichting voor Fundamenteel Onderzoek der Materie} with financial support from the {\it Nederlandse Organisatie voor Wetenschappelijk Onderzoek}, a NATO Collaborative Linkage Grant, and the U. S. Department of Energy Nuclear Physics Division under Contract No. W-31-109-Eng38.


\begin{thebibliography}{99}
\itemsep=0cm
\bibitem{jose98}J. Jos\'e and M. Hernanz, Astrophys. J. {\bf 494}, 680 (1998).
\bibitem{wiescher99}M. Wiescher {\it et al.}, J. Phys. G {\bf 25}, R133 (1999).
\bibitem{audi95}G. Audi and A.H. Wapstra, Nucl. Phys. {\bf A595}, 409 (1995).
\bibitem{magnus90}P.V. Magnus {\it et al.}, Nucl. Phys. {\bf A506}, 332 (1990).
\bibitem{laird02}A.M. Laird {\it et al.}, Phys. Rev. C {\bf 66}, 048801 (2002).
\bibitem{langanke86}K. Langanke {\it et al.}, Astrophys. J. {\bf 301}, 629 (1986).
\bibitem{davids03}B. Davids {\it et al.}, Phys. Rev. C {\bf 67}, 012801(R) (2003).
\bibitem{berg95}A.M. van den Berg, Nucl. Instrum. Methods {\bf B99}, 637 (1995).
\bibitem{fortune78}H.T. Fortune {\it et al.}, Phys. Rev. C {\bf 18}, 1563 (1978).
\bibitem{woertche01}H.J. W\"{o}rtche, Nucl. Phys. {\bf A687}, 321c (2001).
\bibitem{leegte92}H.K.W. Leegte {\it et al.}, Nucl. Instrum. Methods {\bf A313}, 260 (1992).
\bibitem{tilley95}D.R. Tilley {\it et al.}, Nucl. Phys. {\bf A595}, 1 (1995).
\bibitem{hackman00}G. Hackman {\it et al.}, Phys. Rev. C {\bf 61}, 052801(R) (2000).
\bibitem{iliadis02}C. Iliadis {\it et al.}, Astrophys. J. Suppl. Ser. {\bf 142}, 105 (2002).
\end{thebibliography}
\end{document}